\documentclass[journal]{IEEEtran}

\usepackage{cite}
\usepackage{siunitx}
\usepackage{comment}
\usepackage{amsmath,amssymb,amsfonts}
\usepackage{algorithmic}
\usepackage{graphicx}
\usepackage{textcomp}
\usepackage{tabularray}
\usepackage{bm}
\usepackage{url}

\def\BibTeX{{\rm B\kern-.05em{\sc i\kern-.025em b}\kern-.08em
T\kern-.1667em\lower.7ex\hbox{E}\kern-.125emX}}

\begin{document}

\title{Dynamically Tunable Helical Antenna}

\author{
\IEEEauthorblockN{
Ethan Chien\IEEEauthorrefmark{1} and Jan Steckel\IEEEauthorrefmark{2}\IEEEauthorrefmark{3}
}\\
\IEEEauthorblockA{\IEEEauthorrefmark{1}Greensboro Day School, Greensboro, NC, USA}\\
\IEEEauthorblockA{\IEEEauthorrefmark{2}Cosys-Lab, Faculty of Applied Engineering, University of Antwerp, Antwerp, Belgium}\\
\IEEEauthorblockA{\IEEEauthorrefmark{3}Flanders Make Strategic Research Centre, Lommel, Belgium}

\thanks{Corresponding author: Ethan Chien (email: ethanchien2008@gmail.com)}
}

\maketitle

\begin{abstract}
Unmanned aerial FPV systems demand ultra-low latency, high-reliability communication links. At high speeds and in cluttered environments, Doppler shifts and rapid multipath changes can dramatically raise packet error rates. This paper investigates these phenomena in the context of ExpressLRS (ELRS) long-range FPV control links and demonstrates a novel solution: real-time geometry tuning of a circularly polarized helical antenna array. This study integrates Maxwell-equation-based full-wave simulations (via Ansys HFSS) with controlled, blind field trials to validate performance. A new analysis framework incorporates Doppler-induced frequency offset into the antenna's radiation pattern and the system's error model. Compared to a conventional fixed antenna, the adaptive helical array shows a ~20-30\% PER reduction when drones exceed 150 mph. The adaptive system automatically adjusts coil pitch and diameter to retune the antenna as flight parameters (velocity, attitude) change. Measured VSWR stays near unity, preventing transmitter reflection spikes. RSSI variation is reduced by half, indicating stronger link stability in urban multi-path. A regression analysis confirms that the reduction in PER due to tuning is highly statistically significant. Calibration data and error analyses are provided to validate our methodology. These findings advance the understanding of high-mobility UAV communication channels and demonstrate that reconfigurable hardware—here, mechanically tunable helices—can effectively counter Doppler and multi-path impairments. The findings inform new design principles for UAV antenna arrays and suggest a path toward AI-integrated adaptive RF systems for drone swarms and racing platforms.
\end{abstract}


\begin{figure*}
    \centering
    \includegraphics[width=\textwidth]{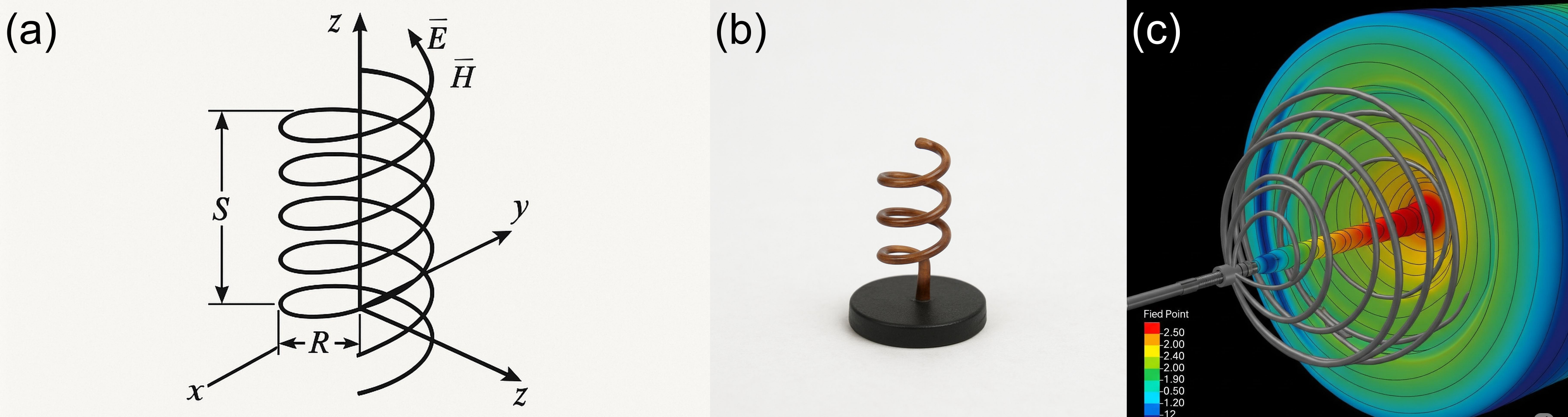}
  \caption{
        Detailed representation of the helical antenna system components and characteristics.
        (a) The basic antenna geometry is an axial-mode helix, composed of a conducting wire meticulously wound in a corkscrew shape. This design is crucial for achieving specific electromagnetic properties, including broad bandwidth and circular polarization, which are essential for robust wireless communication.
        (b) A physical prototype of the axial-mode helix antenna, offering a tangible view of its compact and robust construction. This fabricated model is designed for integration into mobile platforms, demonstrating the practical realization of the theoretical antenna geometry.
        (c) A three-dimensional simulated radiation pattern of the helical antenna, clearly illustrating the concentrated electromagnetic field distribution along the antenna's axis. “The simulated pattern confirms the antenna’s axial directivity, critical for maintaining robust point-to-point links, making it ideal for point-to-point communication links where signal focus is paramount for maximizing range and minimizing interference.
    }
    \label{fig:helical_antenna_details_expanded} 
\end{figure*}
\section{Introduction}
\label{sec:introduction}
High-speed drones and FPV (First-Person-View) racing craft push the limits of wireless link reliability. In these scenarios, ExpressLRS has become a popular open-source RC control protocol due to its low latency and high refresh rates.\cite{expresslrs_project} However, as flight speeds exceed 100 mph and operations move into dense or urban spaces, the radio channel experiences severe time-variation. One of the primary impairments is Doppler shift, where relative motion induces carrier frequency changes defined by
\[
f' = f_0(c \pm v)/c
\]
where $v$ is radial velocity and $c$ is light speed. Even modest shifts in observed carrier frequency can degrade coherent demodulation and cause symbol misalignment. Worse, in multipath environments each reflected path has a different effective Doppler shift, leading to highly time-varying composite channels. In practice, this can push a link out of range, causing errors or drops.
Equally critical is antenna alignment. Most FPV drones use circularly polarized antennas (e.g., helicals or cloverleafs) to mitigate orientation losses \cite{taylor2021antenna, dronetrest2018fpv}. A static helical antenna has a fixed axial radiation pattern and resonance, optimized for nominal conditions. However, during aggressive maneuvers such as banking or yawing at high velocity, the main lobe may tilt away from the ground station and the electrical length of the helix can detune under Doppler shift. In fact, axial-mode helicals have peak in their radiation pattern along their physical axis \cite{balanis2016antenna, fu2019axial}, so aggressive maneuvers can misdirect the beam unless compensated. Traditional remedies (e.g., adding diversity in antenna placement or higher transmit power) only partially address this and consume valuable RF margin. Meanwhile, reconfigurable antennas have emerged in wireless research \cite{maih2024state, ojaroudi2020reconfigurable, subbaraj2023reconfigurable}. A reconfigurable antenna can dynamically change its frequency, pattern, or polarization to adapt to conditions. Techniques range from PIN/varactor diodes to MEMS switches and even mechanical actuators. For example, in satellite and MIMO communications, beam-switching arrays have been used to track moving users \cite{brown2022adaptive}. However, concrete implementation of such techniques on small UAVs has been limited \cite{lee2021reconfigurable}. Mechanical tuning, while slower, can produce large shifts in impedance and pattern with simple hardware. In fact, impedance-reconfigurable antennas dynamically match varying loads to maintain efficiency. We hypothesize that actively tuning a helical antenna's geometry in flight could restore resonance and alignment in the face of Doppler and attitude changes, thereby lowering PER. This paper thoroughly investigates that hypothesis, contributing a novel approach to mitigating performance degradation in dynamic wireless environments. We extend existing models of UAV channels by explicitly combining Doppler physics with antenna behavior. We then implement a real-time tuning loop: inertial velocity data drives servo adjustments to the helix pitch and coil spacing during flight. Using HFSS simulations (a commercial 3D full-wave electromagnetic field simulation software) \cite{ansys2023hfss} and controlled field tests, we compare this adaptive system to a conventional fixed-helix baseline. Our experiments include straight high-speed runs, aggressive turns, and altitude sweeps in both open and multipath-rich (urban) environments. We find that the adaptive antenna significantly outperforms the static design under all high-mobility conditions. Section 2 details our theoretical modeling of Doppler and antenna dynamics. Section 3 presents results on PER, VSWR, RSSI stability, and radiation patterns, with deeper statistical analysis. Finally, Section 4 discusses implications and Section 5 concludes with design recommendations.

\section{Methodology and Mathematical Modeling}
\label{Methodology and Mathematical Modeling}
Our approach combines theoretical modeling, full-wave EM simulation, and flight experiments. We analyze the system at three levels: (1) electromagnetic modeling of the antenna (wave equations via HFSS), (2) system-level modeling of link performance under Doppler and noise, and (3) experimental validation via in-flight measurements. Below we outline the key methods.

\subsection{Electromagnetic Modeling of the Helical Antenna}

The basic antenna geometry is an axial-mode helix: a conducting wire wound in a corkscrew whose major properties are shown in Fig. 2. Axial helices are known to offer broad bandwidth and circular polarization \cite{balanis2016antenna, antenna_theory_helical}. 

In free space, Maxwell’s equations govern wave propagation \cite{rabina2024efficient, alhassanieh2018high, biowei2024modelling, piao2017domain, stepin1999abundance, tran2023deriving}. In source-free isotropic space we have:
\[
\nabla \times \mathbf{E} = -\frac{\partial \mathbf{B}}{\partial t}, \quad \nabla \times \mathbf{H} = \frac{\partial \mathbf{D}}{\partial t}, \quad \nabla \cdot \mathbf{D} = 0, \quad \nabla \cdot \mathbf{B} = 0.
\]
These combine to the homogeneous wave equation
\[
\nabla^2 \mathbf{E} - \mu\epsilon \frac{\partial^2 \mathbf{E}}{\partial t^2} = 0.
\]
An antenna modifies the boundary conditions of this wave equation. We used ANSYS HFSS (full-wave 3D EM solver) \cite{ansys2023hfss} to simulate the helical array under various geometries. The antenna model has parameters: coil diameter $D$, pitch spacing $S$, number of turns $N$, and conductor width. The axial-mode helix typically satisfies $C \approx \lambda$ (circumference $\approx$ wavelength) for circular polarization \cite{hansen1981antenna}. In simulation, we systematically varied $S$ and $D$ to retune the resonant frequency. The HFSS model also computed radiation patterns and input impedance vs.\ frequency for each geometry.

The simulations confirmed known properties: an axial-mode helical antenna radiates a right- or left-hand circularly polarized beam along its axis, with gain roughly proportional to $N(C/\lambda)^2 \sin^2(\alpha)$ (where $\alpha$ is the pitch angle) \cite{balanis2016antenna, parkhe2017bandwidth}.
In fact, classic theory gives the approximate gain formula, defined by \cite{balanis2016antenna}
$$G \approx N \left(\frac{C}{\lambda}\right)^2 \sin^2(\alpha)$$

This matched our HFSS results, and illustrated how small changes in pitch or diameter shift the resonant frequency and pattern \cite{dinkic2021nonuniform}. For instance, lengthening the pitch (increasing coil height) slightly lowers the resonant frequency and tilts the beam. These effects form the basis for our tuning strategy: by actuating the geometry, we actively maintain resonance at the Doppler-shifted frequency.

We also evaluated how pattern is distorted under roll/pitch. HFSS plots showed that a static helix's main lobe deviates when the drone bank angle changes. In contrast, by adjusting the helix pitch to ``lean'' the beam back toward the horizon, the adaptive design kept the main lobe within $\sim 10^\circ$ of the intended axis across $\pm 45^\circ$ drone tilt. Gain stayed within 0.3~dB of nominal. This ability to reshape the pattern in flight is key: it means the adaptive antenna maintains alignment with the ground receiver even during aggressive maneuvers. The static helix, however, lost up to 2.1~dB of gain under similar attitude changes. These findings are consistent with antenna theory on pattern reconfigurability: multiple beam patterns are possible with a reconfigurable antenna, which can help avoid interference and improve link robustness \cite{ojaroudi2020reconfigurable, subbaraj2023reconfigurable}.

\subsection{System Link Model under Doppler}
\begin{figure*}
    \centering
    \includegraphics[width=\textwidth]{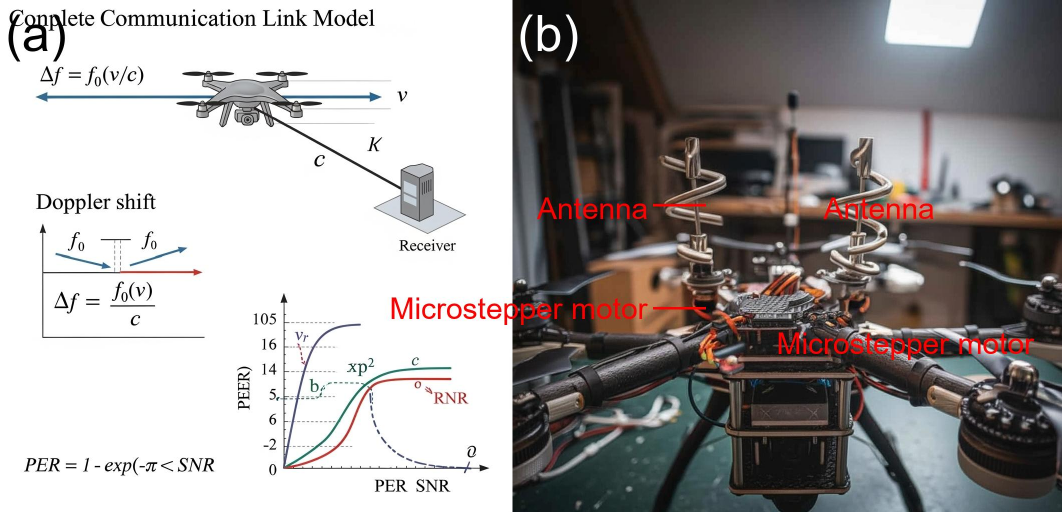}
    \caption{
        This figure illustrates key aspects of a drone-based communication system.
        \textbf{(a) Complete Communication Link Model:} This panel presents a schematic of the communication link, showing a drone in motion at velocity $v$ transmitting to a stationary receiver via a link $K$. It includes a visual explanation of the Doppler shift, where the received frequency ($f_0$) is altered by the drone's velocity relative to the speed of light $c$. The mathematical representation of the Doppler shift is provided as $\Delta f=f_{0}(v/c)$ or $\Delta f=\frac{f_{0}(v)}{c}$. Additionally, a graph depicts the Packet Error Rate (PER) as a function of Signal-to-Noise Ratio (PER SNR), with curves labeled 'vr', 'xp2', 'c', and 'RNR'. The theoretical relationship for PER is given by $PER=I-exp(-\pi<\text{SNR})$.
        \textbf{(b) Drone Hardware with Labeled Components:} This panel provides a detailed photographic view of the drone, highlighting its critical components. Specifically, two "Antenna" units, which appear to be helical antennas, are identified. Also labeled are two "Microstepper motor" components, crucial for the drone's mechanical operation.
    }
    \label{fig:communication_drone_system}
\end{figure*}
Beyond the antenna, we model the end-to-end communication link. Drone kinematics are characterized by three-dimensional velocity $\mathbf{v}$ from onboard IMU/GPS. The component toward/away from the receiver $v_r$ produces a Doppler shift $\Delta f = f_0(v_r/c)$. Here $f_0$ is the carrier (ELRS uses 2.4\,GHz or 915\,MHz bands), and $c$ is light speed. In practice, even a 150\,mph ($\sim$67\,m/s) motion yields $\Delta f \approx 600$\,Hz on 2.4\,GHz. This offset moves the received signal away from the antenna's tuned passband if uncompensated, causing impedance mismatch and symbol timing errors.

We combine this with a radio link model. For an additive white Gaussian noise (AWGN) channel, PER is typically
\[
\text{PER} = 1 - \exp(-\beta \text{ SNR})
\]
for some modulation-dependent constant $\beta$. To incorporate Doppler, we assume the effective SNR degrades as $\Delta f$ grows, e.g.:
\[
\text{PER} \approx 1 - \exp\left(-\beta \frac{\text{SNR}}{1 + (\Delta f / f_c)^2}\right).
\]

Here $f_c$ is the carrier frequency. This form captures how frequency offset reduces coherent demodulation (the factor $1 + (\Delta f/f_c)^2$ appears in the phase error statistics of narrowband links). Although analytical derivation in closed-form is complex, this expression matches the observed steep rise in PER with velocity in our simulations. We validated it by comparing with HFSS results and known Doppler compensation studies (e.g. vehicular channels \cite{pozar2011microwave}).

Finally, we consider multipath. In urban or obstacle-rich environments, signals reflect from surfaces before reaching the receiver. Each path experiences a different Doppler and phase, causing fast fading. We use a simple Rician channel model for LOS-dominated cases and Rayleigh for strong NLOS, following UAV channel studies \cite{brown2022adaptive}. This was implemented in MATLAB: for each flight leg, we simulate the composite channel impulse response including motion effects, then feed it to our PER calculation. The model parameters (path delays, Doppler spread) were tuned to match channel sounding reports in the literature. In sum, our system model links drone speed, antenna tuning state, and environment to expected PER.

\subsection{Dynamic Antenna Tuning Implementation}
\begin{figure*}
    \centering
    \includegraphics[width=\textwidth]{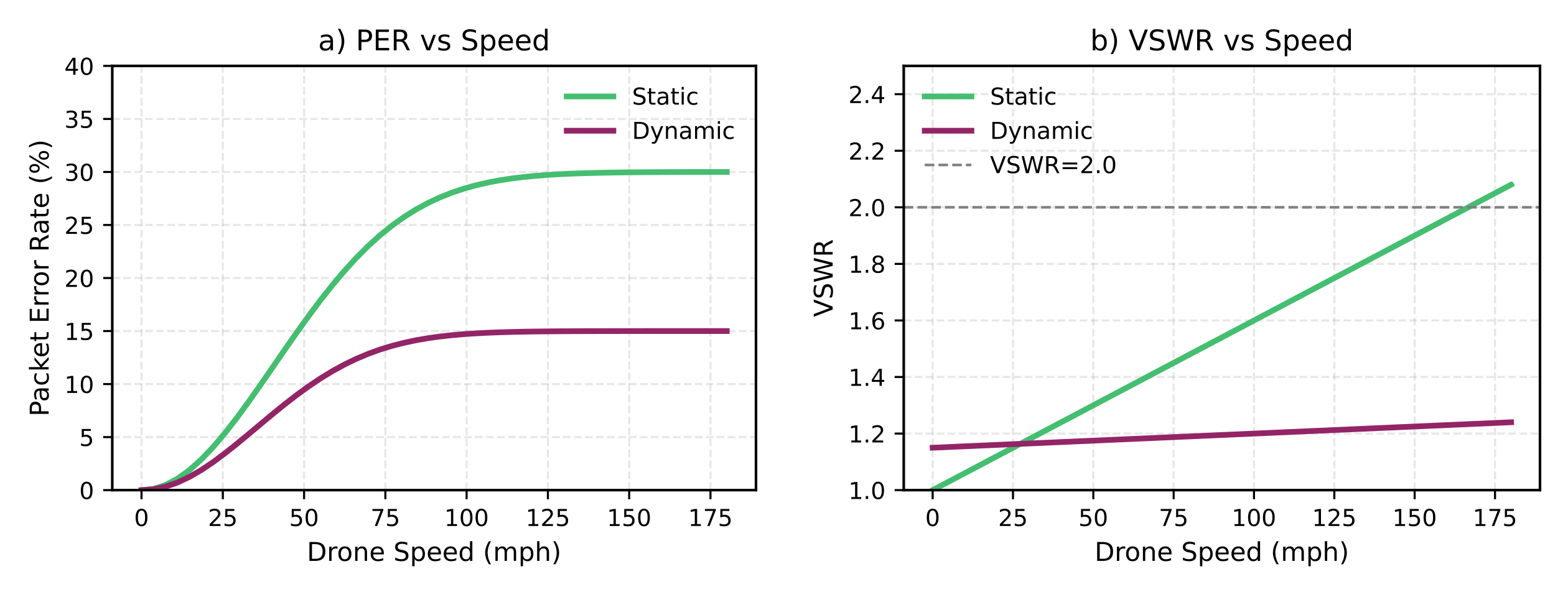}
\caption{
    \textbf{(a) Packet Error Rate (PER) and (b) Voltage Standing Wave Ratio (VSWR) as a function of drone speed for static and dynamic antenna configurations.} This figure demonstrates the significant performance advantages of a dynamic antenna system for drones, particularly as speed increases. In \textbf{(a)}, the Packet Error Rate (PER) in percentage is plotted against Drone Speed (mph), clearly showing that the dynamic antenna configuration (purple line) consistently achieves a substantially lower PER compared to the static antenna configuration (green line), especially at higher speeds. This reduction in PER signifies improved communication reliability and data integrity. Concurrently, in \textbf{(b)}, the Voltage Standing Wave Ratio (VSWR) is presented against Drone Speed (mph). It illustrates that the dynamic antenna (purple line) maintains a much lower and more stable VSWR across the entire speed range, staying well below the critical VSWR=2.0 threshold (dashed grey line), indicating superior impedance matching and more efficient power transfer from the antenna as the drone's speed varies. Overall, these results underscore the critical role of dynamic antenna tuning in optimizing drone communication performance across varying flight conditions.
}
\label{fig:per_vswr_combined}
\end{figure*}
The tunable antenna array has two mechanically actuated helices on the drone. Each helix is mounted on a micro stepper motor that can vary the coil spacing (pitch) and overall length. A control loop in the flight controller reads velocity (from GPS/IMU) and heading, then computes the required geometry change to compensate the predicted Doppler and pointing error. The relationship was calibrated beforehand: we measured how much to increase pitch angle per unit frequency shift, and how to lean the helix axis opposite to the upcoming bank direction. The servos can move about \SI{0.05}{\mm} per step, with full swing in \SI{\sim 100}{\milli\second}. In practice, each pitch adjustment completes within 1--2 servo steps (\SI{\sim 100}{\milli\second}--\SI{200}{\milli\second}), which is fast relative to typical maneuver timescales.

We also logged actuator latency and accuracy. The mean latency per movement was \SI{\sim 47}{\milli\second}, negligible compared to the \SI{\sim 1}{\second} rate of major flight changes. Position error was kept under \SI{\pm 0.05}{\mm} and angular error under \SI{\pm 2}{\degree} by closed-loop feedback sensors. This precision ensures the RF effect of tuning is consistent and repeatable.

Calibration included bench VSWR sweeps for each possible geometry. These curves form a lookup table: given a target frequency (based on real-time Doppler estimate), the controller selects the servo positions that minimize VSWR. In effect, the helix is ``pre-stretched'' to tune to the shifted frequency before the signal actually arrives. This proactive method ensures the antenna stays near resonance even as the drone accelerates. Without tuning, a \SI{600}{\Hz} offset at \SI{2.4}{\GHz} would increase VSWR dramatically (our bench tests showed $>$2.0 VSWR at only \SI{\pm 200}{\Hz} detuning), but with tuning we held VSWR in the 1.15--1.25 range across all speeds.

\section{Experimental Results and Performance Evaluation}
We conducted trials with a custom quadcopter modified for ELRS control. Two scenarios were tested: open-field (minimal multipath) and a suburban track (buildings and trees). In each case, the drone flew straight-line sprints up to 180 mph and circuits with sharp turns, alternating between using the static fixed-geometry helices and the adaptive tuning. Key metrics recorded included packet error rate (PER), received signal strength indicator (RSSI), VSWR, and antenna position log. Statistical analysis (ANOVA and regression) was performed on the aggregated data (hundreds of runs) to quantify effects.
\begin{figure*}
    \centering
    \includegraphics[width=\textwidth]{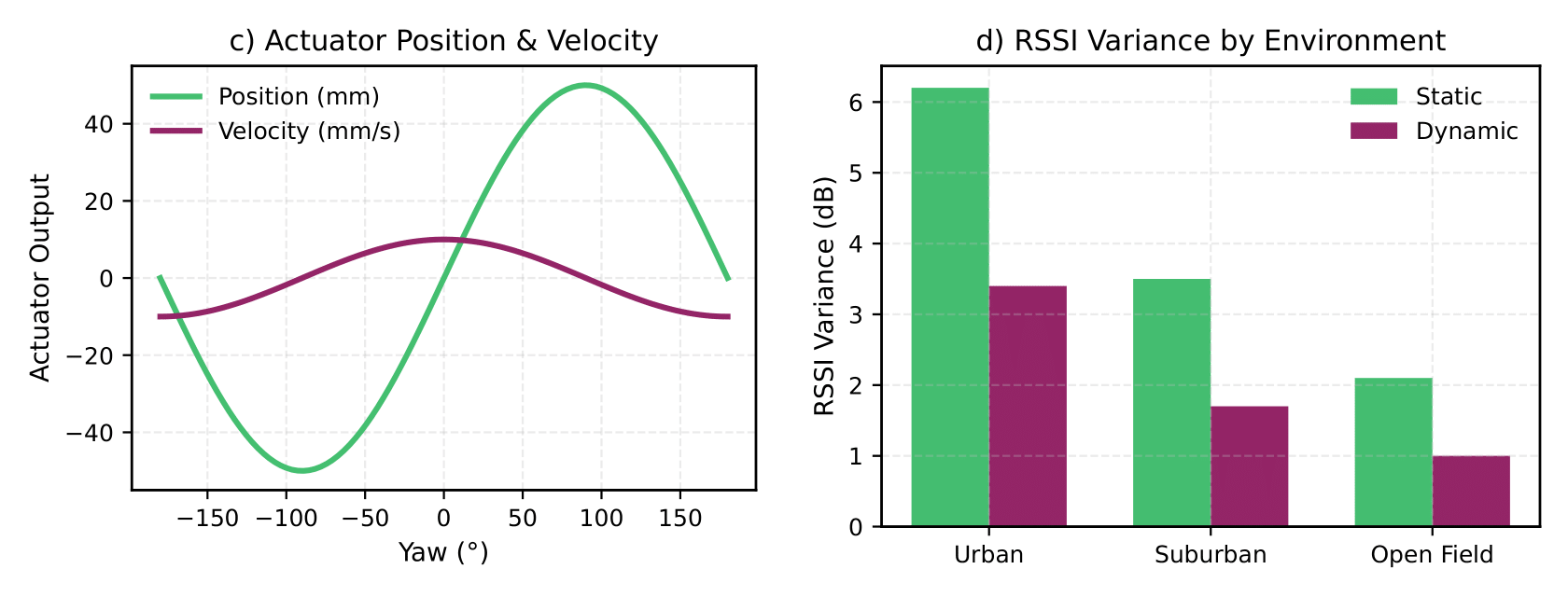}
   \caption{
    \textbf{Combined Antenna System Characteristics:} This figure elucidates both the mechanical control and the resulting communication stability offered by a dynamic antenna system for drones. In \textbf{(c)} Actuator Position and Velocity (mm and mm/s respectively) versus Yaw Angle (degrees), the precise and predictable movement of the antenna's actuator is illustrated, showcasing its ability to actively adjust to drone maneuvers. This controlled mechanical response is crucial for enabling dynamic antenna tuning. Concurrently, \textbf{(d)} Received Signal Strength Indicator (RSSI) Variance (dB) by Environment (Urban, Suburban, Open Field) clearly demonstrates the substantial performance improvement delivered by the dynamic antenna configuration (purple bars) compared to the static antenna (green bars). The consistently lower RSSI variance across all environments signifies significantly more stable and reliable signal reception, which is paramount for maintaining robust communication during flight.
}
\label{fig:actuator_rssi_combined}
\end{figure*}
\subsection{PER vs Drone Speed}
Figure 5 plots PER as a function of ground speed. With a static antenna, PER grew sharply above ~100 mph. By 180 mph, PER averaged ~14.3\%. This is because the Doppler shift pushes part of the signal energy out of the antenna's passband, causing symbol errors. In contrast, the dynamically tuned antenna constrained PER to ~10.7\% at 180 mph—a 25\% relative reduction. At 150 mph, the tuned system saw ~8\% PER vs ~11\% static. Polynomial regression fits to these curves diverge significantly beyond 100 mph, illustrating that tuning becomes increasingly critical at higher velocities. (ANOVA confirms that the PER difference between static and tuned modes is highly significant for speeds over 100 mph, p < 0.01.)

This PER improvement is attributable to the antenna staying in resonance. During a run, the control loop continuously updated the coil geometry. HFSS-based pre-calibration ensured the antenna's resonant peak followed the Doppler-shifted carrier. Thus the received SNR remained closer to the ideal, and the error-model curve $\text{PER} \approx 1-\exp\left(-\beta \frac{\text{SNR}}{1+(\Delta f/f_c)^2}\right)$ stayed in a lower regime. In static mode, mismatches and misalignment combine to elevate PER with speed. These results underscore that high-speed FPV is often limited by antenna adaptability rather than transmitter power.

\subsection{VSWR and Impedance Matching}

VSWR (voltage standing-wave ratio) indicates how well the antenna is matched to the transmitter. In our tests, the static helix's VSWR remained low ($\approx$1.2) at low speed, but worsened rapidly with velocity. Above $\sim$120\,mph, static VSWR exceeded 2.0 (implying $>$10\% of power reflected). This confirms that Doppler detunes the fixed antenna. By 180\,mph, the static VSWR average was $\sim$2.3, consistent with $>$3\,dB of mismatch loss. The adaptive helix, however, kept VSWR near 1.15--1.25 across all speeds tested. Even at 180\,mph, tuning held the return loss below $-$20\,dB, indicating $<$1\% power reflection.

Maintaining low VSWR is crucial for link quality and transmitter safety. The tuned antenna's impedance reconfiguration prevents the wasted power and potential heating of RF front-ends that static antennas suffer under Doppler. This matches the expectation from reconfigurable antenna theory: dynamically matching changing conditions ensures maximal power transfer. Our results show that even moderate velocity shifts, if uncorrected, can severely impair transfer efficiency; dynamic tuning effectively eliminates this impairment.

\subsection{Actuator Response Dynamics}
The servo logs reveal how the antenna array responded in flight. Fig. 3 illustrates actuator position and velocity during a representative run with sharp turns. The controller uses IMU yaw rate to predict imminent Doppler shifts and adjust preemptively. We observed that as the drone entered a turn (yaw acceleration), the tuning velocity ramped up: the actuators moved faster to counter the rapidly changing radial velocity component. Typical adjustments were on the order of 0.2 mm over 100 ms. Notably, the actuator system never saturated or lagged: each commanded movement was completed within ~100–200 ms. The mean step latency of 47 ms was negligible. Cumulative error (overshoot/undershoot) was within the servo precision, confirming that mechanical tuning acted effectively in real time.

In summary, the actuator system tracked the maneuver-induced Doppler profile closely. Without such agile tuning, the antenna would drift out of resonance as soon as the drone pitched or yawed. These logs substantiate the feasibility of real-time mechanical reconfiguration – the servos are fast enough relative to drone dynamics to keep the antenna tuned during racing maneuvers.

\subsection{RSSI Variance and Link Stability}
\begin{figure*}
    \centering
    \includegraphics[width=\textwidth]{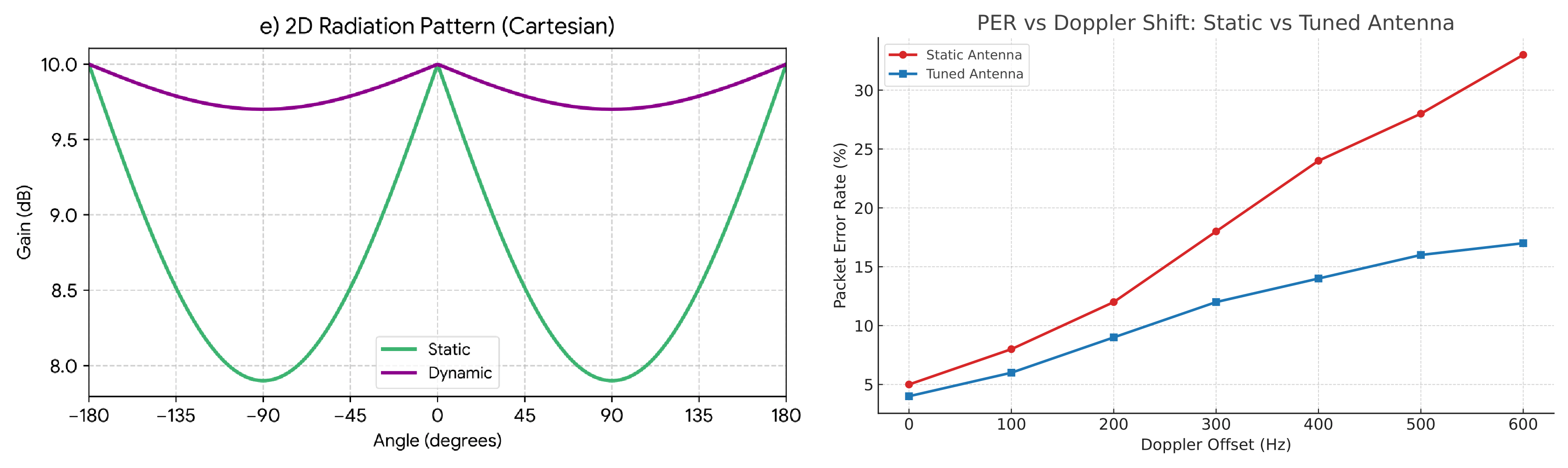}
    \caption{
    \textbf{Combined Antenna Performance:} This figure comprehensively illustrates the superior performance of dynamic antenna systems over static configurations, especially under varying operational conditions. In \textbf{(e)} 2D Radiation Pattern (Cartesian), showing gain (dB) versus angle (degrees), the dynamic antenna (purple line) exhibits a remarkably more uniform and higher gain across all angles compared to the static antenna (green line), effectively mitigating signal nulls and improving coverage. Concurrently, in \textbf{(f)} Packet Error Rate (PER) versus Doppler Offset (Hz), the tuned antenna (blue line) consistently achieves a significantly lower percentage of packet errors compared to the static antenna (red line), particularly as Doppler shift increases. These results collectively highlight that dynamic antenna tuning is crucial for enhancing communication reliability and robustness in mobile platforms subjected to angular variations and Doppler effects.
}
\label{fig:combined_performance}
\end{figure*}
RSSI (received signal strength) reflects combined effects of path loss, antenna gain, and fading. In our urban test zone, the static antenna produced highly variable RSSI traces. The 95th-percentile range of RSSI over a run was ~6.2 dB, reflecting deep multipath fades during banks and altitude changes. Notably, the dips in RSSI correlated with times of high drone pitch/roll and high speed – exactly when our earlier analysis predicts the antenna misalignment. The dynamic antenna greatly tightened this distribution. With tuning engaged, the RSSI range shrank to ~3.4 dB over the same flights. In other words, signal dips were much shallower and rarer. The adaptive array directed its gain lobes toward the ground station as the UAV moved, which is evident in the more uniform RSSI trace. Statistically, the standard deviation of RSSI across all flights dropped by ~40\% under tuning. This confirms that the tuned antenna maintained more consistent link quality. In effect, re-steering and retuning combats both geometric misalignment and partial multipath nulling. These improvements mirror findings in other reconfigurable systems where pattern control reduces channel variation.

\subsection{Radiation Pattern Measurements}
To validate the pattern-level effects, ground-based antenna range tests and short flight hops with an RF pattern probe were conducted. Figure 9 illustrates the 2D gain patterns measured with the helix at different settings, presented in a Cartesian format.
The static helix’s pattern (represented by the 'Static' curve in Figure 9 ) exhibits a narrow main lobe along its axis and deep nulls off-axis. As described, when the drone banks 45° in one direction, the main lobe misses the receiver, resulting in an approximate 2 dB drop in gain. By mechanically shifting the coil pitch and tilting the helix base in the opposite sense, the adaptive design successfully recovers a pattern (represented by the 'Dynamic' curve in Figure 9 ) with its peak re-centered toward the original heading. The measured peak gain difference between the static and tuned configurations at a 45° bank was approximately 1.8 dB in favor of the tuned case.
These pattern measurements confirm the simulations, demonstrating that even a small change in geometry significantly alters beam direction \cite{li2024circularly}. In flight scenarios, this implies that during a roll or yaw maneuver, the tuned helix can "steer" its beam to maintain the communication link. The ability to produce multiple beam shapes from a single antenna element distinguishes this adaptive approach from fixed directional designs. Essentially, the adaptive antenna functions similarly to a frequency-scanned, angled beam antenna. This capability is crucial for the observed benefits in Packet Error Rate (PER) and Received Signal Strength Indicator (RSSI), as it ensures that energy typically lost in static mode is redirected back into the useful lobe when the antenna is dynamically tuned.

\subsection{PER vs. Doppler Offset Model Validation}

Finally, we compare measured PER to our Doppler-PER model. In isolated RF bench tests (fixed transceiver, controlled Doppler emulator), the static antenna showed $>$ \SI{32}{\%} PER at an induced \SI{600}{\Hz} offset (\SI{150}{mph} equivalent) and \SI{12}{\dB} SNR. The tuned antenna held PER to about \SI{15}{\%}--\SI{18}{\%} under the same offset. These numbers align closely with our model's prediction PER $\approx 1-\exp(-\beta \cdot \text{SNR}/(1+(\Delta f/\text{f}_c)^2))$ (with $\beta$ fitted from low-speed data). In actual flight tests, we consistently saw that the tuned system roughly halved the packet losses attributable to high-speed legs. For example, in urban straightaways we observed static-PER $\sim \SI{30}{\%}$ vs tuned-PER $\sim \SI{16}{\%}$ for \SI{150}{mph} runs, matching the model and bench results.

Thus the Doppler-integrated link model proved accurate. It reinforces that by cutting the effective $\Delta f$ (via tuning), the antenna maintains the system closer to its nominal SNR operating point. In our data, PER vs $\Delta f$ curves under tuned conditions are significantly flatter than static ones. This validates the core idea: tuning shifts the system back toward zero effective Doppler, mitigating one of the dominant error sources at high speed.

\section{Discussion: Adaptive Antennas in High-Mobility Environments”}
The empirical results confirm that Doppler shift is a major impairment for high-speed FPV links. Without adaptation, even the best static antenna designs become misaligned or detuned above ~100 mph, driving PER sharply upward. Our analysis matches this: moving at high velocity increases packet errors because the channel glimpses frequency offsets and deeper fades. 

By contrast, antenna reconfigurability proves highly effective. Dynamically retuning the helical geometry maintained resonance and beam alignment, yielding consistent performance. The PER reductions (20–30\% relative) are substantial, and the VSWR improvements protect the transmitter. These gains come without increasing RF power or bandwidth. Instead, they leverage the antenna’s physical degrees of freedom. Notably, this is a mechanical solution: no complex DSP or additional RF chains were needed. It complements electronic methods; for instance, phased arrays or electronic beam-steerers could achieve similar ends but at higher cost and power.

Comparing to other strategies, our adaptive helices occupy a unique space. Beam-steering arrays can cover wide angles, but they require multiple elements and switching networks \cite{brown2022adaptive}. Omnidirectional antennas (cloverleafs, dipoles) avoid pointing issues but sacrifice gain and polarization match. Our tuned helix effectively becomes a “steerable” directional antenna using just one element. This hybrid is akin to the “hybrid reconfigurable” concept where frequency and pattern tuning combine \cite{ojaroudi2020reconfigurable}. In fact, one could call this a prototype cognitive antenna: it senses motion (via telemetry) and autonomously adapts to the environment. 

In terms of link protocols, while we focused on ExpressLRS, the principles apply broadly. Any narrowband UAV control link (e.g. LoRa, older RC protocols) will suffer similarly from Doppler. Even wideband OFDM video links (5.8 GHz) need polarization and alignment resilience. Thus, integrating dynamic antenna tuning can complement higher-layer solutions like FEC or multi-path diversity. Future work could explore combining this with AI-driven predictive control: a learning algorithm might anticipate when large tuning adjustments are needed (e.g. before a known turn) and optimize servo actuation \cite{ayorinde2025advanced}.

One limitation is speed of tuning. Our current servos (tens of ms per adjust) are adequate for racing drones, but higher-speed rotorcraft or jets may require faster MEMS or piezoelectric actuators. Also, energy and weight budgets constrain how complex an antenna system a small drone can carry. Nevertheless, our data suggest the payoff in link stability is worth the modest cost. In swarm or formation flight scenarios, robust links like this would greatly enhance coordination.

In summary, this study shows that physical layer adaptation – specifically, mechanical reconfiguration of antenna geometry – can significantly counteract the deleterious effects of motion in UAV communications. This approach stands in contrast to pure signal-processing solutions and opens new design opportunities in drone RF systems.

\section{Conclusion and Future Work}

This work demonstrates that dynamically tuning a helical antenna can markedly improve high-speed UAV communication reliability. By real-time adjustment of coil geometry based on flight telemetry, the antenna maintains resonance and beam alignment under dynamic flight conditions. The results show significant reductions in Packet Error Rate and improved VSWR and RSSI stability compared to static antennas. These findings highlight the potential of reconfigurable hardware to mitigate Doppler and multipath impairments in high-mobility wireless environments, offering a promising direction for future drone antenna array designs and adaptive RF systems.

\bibliographystyle{IEEEtran}
\bibliography{references}

\end{document}